\documentclass[aps,prl,twocolumn,showpacs,superscriptaddress,groupedaddress]{revtex4}  
\usepackage{graphicx}  
\usepackage{dcolumn}   
\usepackage{bm}        
\usepackage{amssymb}   

\begin{document}



\title{First dark matter search results from the PandaX-I experiment}
\date{\today}
\vspace{0.5in}
\newcommand{\sjtuphys}{\affiliation{INPAC and Department of Physics and Astronomy, \\Shanghai Jiao Tong University, Shanghai, 200240, P. R. China}}
\newcommand{\sjtume}{\affiliation{School of Mechanical Engineering, \\Shanghai Jiao Tong University, Shanghai, 200240, P. R. China}}
\newcommand{\sdu}{\affiliation{School of Physics and Key Laboratory of Particle Physics and Particle Irradiation (MOE), Shandong University, Jinan 250100, China}}
\newcommand{\sinap}{\affiliation{Shanghai Institute of Applied Physics, Chinese Academy of Sciences, Shanghai, 201800, P. R. China}}
\newcommand{\umich}{\affiliation{Department of Physics, University of Michigan, Ann Arbor, MI, 48109, USA}}
\newcommand{\pku}{\affiliation{School of Physics, Peking University, Beijing, 100080, P. R. China}}
\newcommand{\umd}{\affiliation{Department of Physics, University of Maryland, College Park, MD, 20742, USA}}
\newcommand{\yalong}{\affiliation{Yalong River Hydropower Development Company, Ltd., 288 Shuanglin Road, Chengdu, 610051, P. R. China}}

\sjtuphys
\sjtume
\sdu
\sinap
\umich
\pku
\umd
\yalong

\author{Mengjiao Xiao} \sjtuphys
\author{Xiang Xiao} \sjtuphys
\author{Li Zhao} \sjtuphys 
\author{Xiguang Cao} \sinap
\author{Xun Chen} \sjtuphys
\author{Yunhua Chen} \yalong
\author{Xiangyi Cui} \sjtuphys
\author{Deqing Fang} \sinap
\author{Changbo Fu} \sjtuphys
\author{Karl L. Giboni} \sjtuphys
\author{Haowei Gong} \sjtuphys
\author{Guodong Guo} \sjtuphys
\author{Jie Hu} \sjtuphys
\author{Xingtao Huang} \sdu
\author{Xiangdong Ji} \thanks{Spokesperson: xdji@sjtu.edu.cn and xji@umd.edu} \sjtuphys\pku\umd
\author{Yonglin~Ju} \sjtume
\author{Siao Lei} \sjtuphys
\author{Shaoli Li} \sjtuphys
\author{Qing Lin} \sjtuphys
\author{Huaxuan Liu} \sjtume
\author{Jianglai~Liu} \thanks{Corresponding author: jianglai.liu@sjtu.edu.cn} \sjtuphys
\author{Xiang Liu} \sjtuphys
\author{Wolfgang Lorenzon} \umich
\author{Yugang Ma} \sinap
\author{Yajun Mao} \pku
\author{Kaixuan Ni}  \thanks{Corresponding author: nikx@sjtu.edu.cn} \sjtuphys
\author{Kirill Pushkin} \sjtuphys\umich
\author{Xiangxiang Ren} \sdu
\author{Michael Schubnell} \umich
\author{Manbing Shen} \yalong
\author{Scott Stephenson} \umich
\author{Andi Tan} \umd
\author{Greg~Tarl\'e} \umich
\author{Hongwei Wang} \sinap
\author{Jimin Wang} \yalong
\author{Meng Wang} \sdu
\author{Xuming Wang} \sjtuphys
\author{Zhou Wang} \sjtume
\author{Yuehuan Wei} \thanks{Current institution: University of Z\"urich} \sjtuphys 
\author{Shiyong Wu} \yalong
\author{Pengwei Xie} \sjtuphys
\author{Yinghui You} \yalong
\author{Xionghui Zeng} \yalong
\author{Hua Zhang} \sjtume
\author{Tao Zhang} \sjtuphys
\author{Zhonghua Zhu} \yalong
\collaboration{The PandaX Collaboration}
\
\date{\today}

\begin{abstract}
We report on the first dark-matter (DM) search results from PandaX-I, a low threshold dual-phase xenon experiment operating at the China JinPing Underground Laboratory.  In the 37-kg 
liquid xenon target with 17.4 live-days of exposure, no DM particle candidate 
event was found.  This result sets a stringent limit for low-mass DM particles and disfavors 
the interpretation of previously-reported positive experimental results.  The 
minimum upper limit, $3.7\times10^{-44}$\,cm$^2$, for the spin-independent isoscalar DM-particle-nucleon scattering cross section is obtained at a DM-particle mass of 49\,GeV/c$^2$ 
at 90\% confidence level. 

\end{abstract}

\maketitle

The dark matter is a leading candidate to explain 
gravitational effects observed in galactic rotational curves, galaxy clusters, and
large scale structure formation, etc. ~\cite{dm:evidence}.  Weakly interacting 
massive particles (WIMPs), a particular class of DM candidates, are interesting 
in particle physics and can be studied in colliders, indirect and direct
detection experiments~\cite{dm:expreview}. These particles
can naturally arise from models beyond the Standard Model of particle
physics, such as supersymmetry
and extra dimensions~\cite{Jung:1996}. Direct positive detection of WIMPs using
ultra-low background detectors in deep underground laboratories would provide 
convincing evidence of DM in our solar system and allow the probing of 
fundamental properties of WIMPs.

Direct detection experiments using different technologies have 
produced many interesting results. The first reported positive observation 
was from the DAMA/LIBRA experiment which used NaI(Tl) crystal
as targets~\cite{ref:dama}. The results can be explained by WIMPs with masses around
10 or 50 GeV/$c^2$~\cite{ref:dama_savage}. More recently, experiments using ultra-low
threshold detectors with a germanium target (CoGeNT~\cite{ref:cogent}) and a
silicon target (CDMS II-Si~\cite{ref:cdms-si}) reported signals above
background, pointing to a low-mass WIMP particle near 10\,GeV/$c^2$.
In addition, the CRESST-II experiment using the CaWO$_4$ crystal 
also reported signals indicating 10 or 30 GeV/$c^2$ WIMPs, 
but not confirmed by the upgraded detector~\cite{ref:cresst1,ref:cresst2}. 
These results have produced much excitement in the community~\cite{ref:theoryoverview}
and call for further examinations of the low-mass WIMP signals through
other experiments~\cite{ref:cdex0,ref:cdex,ref:cdmslite,ref:supercdms}.

In recent years, new techniques using noble liquids (xenon, argon)
have shown exceptional potential due to the capability of 
background suppression and discrimination, and scalability to large target masses.
The XENON10/100~\cite{xenon10_0,xenon10, xenon100_first, xenon100_2nd, xenon100_final} and LUX~\cite{ref:lux} experiments using the dual-phase technique have 
improved WIMP detection sensitivity by more than two orders of magnitude 
in a wide mass range.

The PandaX experiment, operated at the China JinPing Underground Laboratory
(CJPL)~\cite{ref:cjpl}, uses the dual-phase xenon technique to search for
both low and high mass WIMP dark matter. The first stage of PandaX
(PandaX-I) employs a pancake-shaped time projection chamber (TPC)
with about 120-kg active xenon target mass. This TPC, designed 
with a high light-yield thus low-energy threshold, is dedicated to searching 
for low-mass DM particles. A detailed description of the PandaX-I experiment 
and CJPL is given in Ref.~\cite{pandax:tdr}.

The PandaX TPC with a diameter of 60\,cm and drift length of 15\,cm 
is mounted in a stainless steel inner vessel, 75\,cm in diameter and 103\,cm in height, containing a total mass of 450\,kg of liquid xenon (LXe).
The inner vessel is over-dimensioned to accommodate the future upgrade to PandaX-II.
It is contained in a vacuum cryostat constructed from 5-cm thick high-purity oxygen-free copper, and enclosed by a passive shield made of copper, polyethylene, lead, and polyethylene, from inner to outer layers.
The gap between the copper cryostat and inner copper shield is continuously flushed with dry nitrogen gas to reduce radon to below 10\,Bq/m$^3$. The cryogenics and gas handling systems are installed outside the shield and maintain the LXe at a working temperature of
179.5\,K and an absolute pressure of about 2.0\,atm. During data taking the cryogenics system 
provided a thermal and pressure stability of better than 1\%. 
Xenon is continuously recirculated through a getter purification system at a rate around 
30 SLPM, resulting in a stable electron lifetime of approximately 260\,$\mu$s. 

The dual-phase xenon TPC technique enables the detection of both the primary
scintillation signal (S1) in the liquid and the ionization signal through proportional
scintillation (S2) in the gas. This allows discrimination of nuclear recoils (NR) from 
electron recoils (ER) via the S2 to S1 ratio~\cite{AprileDoke2010}. Further background reduction is achieved through fiducialization of the target volume using 3D event position reconstruction.
The PandaX-I TPC is operated with cathode, gate, and anode potentials setting to $-$15\,kV, $-$5\,kV, and ground, respectively. 
This generates an expected drift field of 667 V/cm, which agrees within 3\% with the average simulated from the actual geometry. The electron drift velocity is 1.7mm/$\mu$s according to Ref.~\cite{Yoshino:76}, in excellent agreement with drift time distribution in our data.
The liquid level is centered between the gate and anode, which are separated by 8\,mm. The level can be adjusted to a precision of 0.1\,mm with an externally-controlled overflow mechanism.

The scintillation light is collected by two opposing horizontal arrays of photomultiplier
tubes (PMTs). The bottom array, consisting of 37 3-inch Hamamatsu R11410-MOD PMTs,
is immersed in LXe 5\,cm below the cathode. The top array consists of 143 1-inch
Hamamatsu R8520-406 PMTs and is mounted in the gaseous xenon above the
anode. The horizontal position of an event is reconstructed using the S2 signal
captured by the top array, while the vertical position is determined using the time difference
between S1 and S2 signals. During operation, three top PMTs and two bottom PMTs
were disabled due to malfunction.

The PMT gains are adjusted to $~2\times10^6$, calibrated weekly using low-intensity LED light, and are stable within 10\% over time. Random hits are used to monitor the
PMT dark rates and gains. During the DM search run,
dark rates of the top and bottom PMTs are approximately 50 Hz
and 1 kHz, respectively.  These rates have 
correlation with detector parameters such as cryogenic conditions,
TPC high voltage, as well as ambient temperature. The data with spurious dark
rates are removed from this analysis. 

The raw signals from the PMTs get amplified by a factor of 10 through 
Phillips 779 amplifiers, then fed into CAEN V1724 14-bit 100 MS/s digitizers. 
The event trigger is constructed by summing, integrating, and 
discriminating the time-over-threshold (``Majority'') outputs 
from the digitizers for the bottom PMTs.
For events triggered by S2 signals relevant for DM search, the trigger threshold corresponds to 89\,photoelectrons (PEs). Waveforms from the PMTs are recorded 100\,$\mu$s before and after the trigger, with zero length encoding for signals below 1/3\,PE on each channel.

We analyzed waveforms for each PMT channel to define physical events. 
Hits are identified from each PMT waveform with a threshold corresponding to about 40\% single PE amplitude.
These hits are clustered in time to form physical
signals. Consistent selection results were obtained by alternatively implementing signal finding
on the summed waveform.
The S1 and S2 signals are identified primarily using their widths.
An event relevant for DM search contains a single S1 signal before the S2 signal.

Data quality filters are applied to
separate physical signals from noise. The 200 kHz noise originating from
the PMT high-voltage supplies is filtered based on its ringing signature.
Events with large baseline variations as well as abnormal S2 signal ratios of the bottom to top PMTs are rejected. A good S1 signal requires at least two PMTs fired and an appropriate pulse height-to-area ratio.

The horizontal position of an event is reconstructed by both a center-of-gravity (CoG) and 
neural network (NN) algorithm using the S2 pattern observed with the top PMT array. 
The average difference of the positions for the two methods is about 1 cm, and events with 
large difference due to abnormal S2 hit patterns are rejected. In
our analysis, we use the CoG approach, and have confirmed that the 
conclusions do not change when the latter approach is used.

Calibration runs with $^{137}$Cs and $^{60}$Co gamma sources and a $^{252}$Cf neutron source, deployed between the outer and inner vessels, were taken to characterize the detector response. The neutron source produces both elastic NR events as well as inelastic events with gamma energies of 40 ($^{129}$Xe) and 80\,keV ($^{131}$Xe). 
The 40 keV inelastic events are utilized in the uniformity corrections to the S1 and S2 signals. 
After subtracting the NR contribution, the ER light yield is 5.1 (4.7)\,PE/keV$_{\rm ee}$ (electron-equivalent) at 40 (80)\,keV. Extrapolating these light yields using the NEST model~\cite{ref:NEST}, we obtain a light yield at 122\,keV
of 4.2$\pm$0.2\,PE/keV$_{\rm ee}$ at a drift field of 667\,V/cm and 7.3$\pm$0.3\,PE/keV$_{\rm  ee}$
at zero field ($L_y^{122}$).

\begin{figure}[!htbp]
  \centering
  \includegraphics[width=0.49\textwidth]{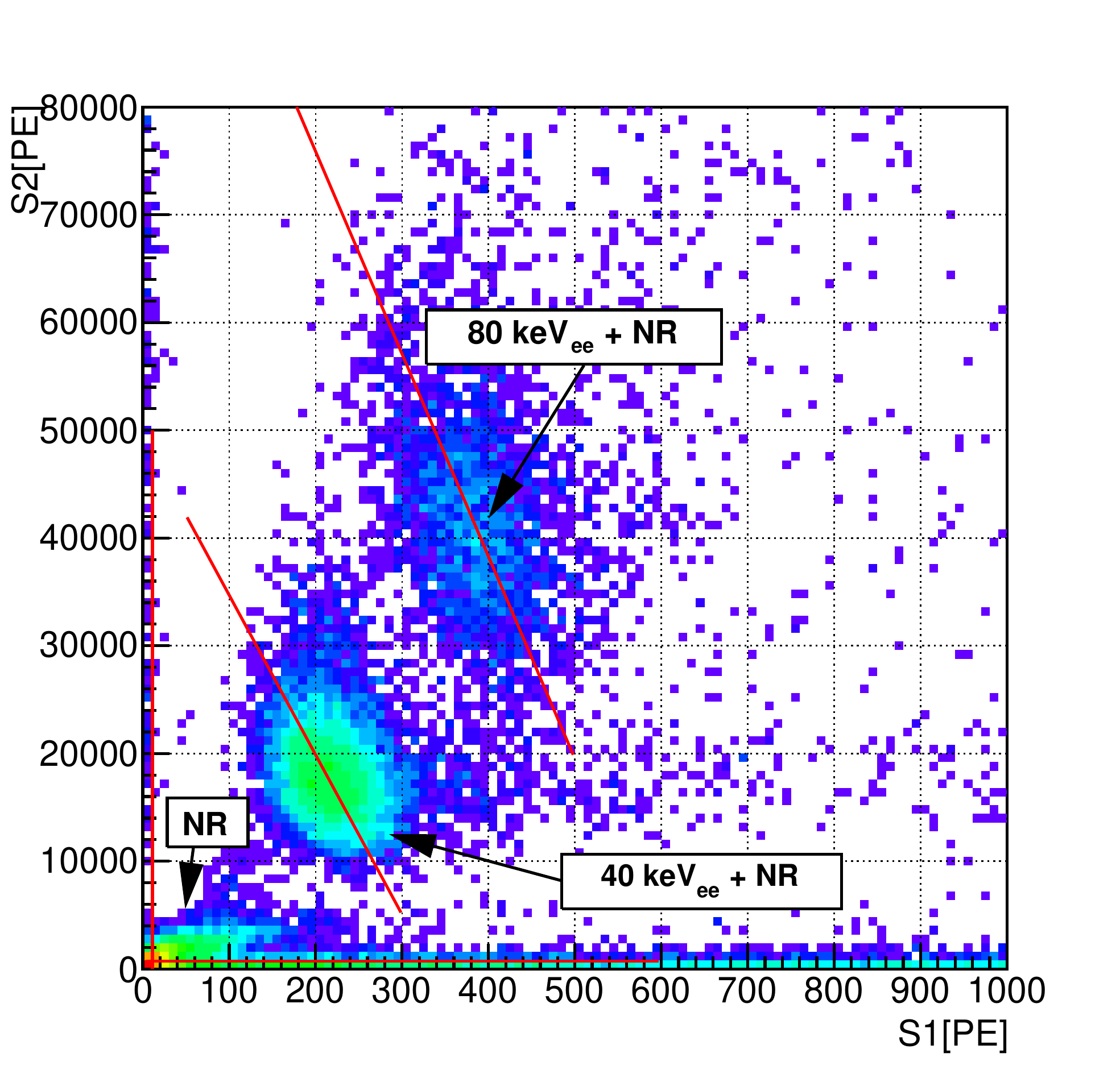}
  \caption{S2 versus S1 distribution from $^{252}$Cf calibration data.
   The horizontal and vertical lines close to the axes indicate the average NR contribution that is
    subtracted from the inelastic peaks when performing the anti-correlation fit, 
	and the off-diagonal lines are the
    fit results.
  }
  \label{fig:ces_40_80}
\end{figure}

The S1--S2 combined energy scale for ERs is defined as ${E_{\rm ee}^{\rm ce} = W \cdot (S1/\alpha + S2/\beta)}$. The work function $W=13.7$\,eV~\cite{ref:NEST} is the average energy needed to produce a quantum (photon or electron) in liquid xenon, $\alpha$ is the photon detection efficiency (PDE), and $\beta$ is the product of the single-electron gas gain and the electron extraction efficiency (EEE).
The single-electron gas gain is determined to be 22.1\,PE/e with a 45\% resolution by fitting single-electron S2 signals. 
The PDE and EEE are found to be (10.5$\pm$0.4)\% and (79.8$\pm$7.0)\%, respectively, by fitting the anti-correlation between S1 and S2 for 40 and 80 keV inelastic events (Fig.~\ref{fig:ces_40_80}). The uncertainties in these values are estimated from the difference between the values extracted from the two gamma peaks.
The scintillation and ionization yields at 40, 80, and 662\,keV ($^{137}$Cs) obtained from our data are consistent with the NEST predictions within 10--20\%.

\begin{figure}[!htbp]
  \centering
  \includegraphics[width=0.49\textwidth]{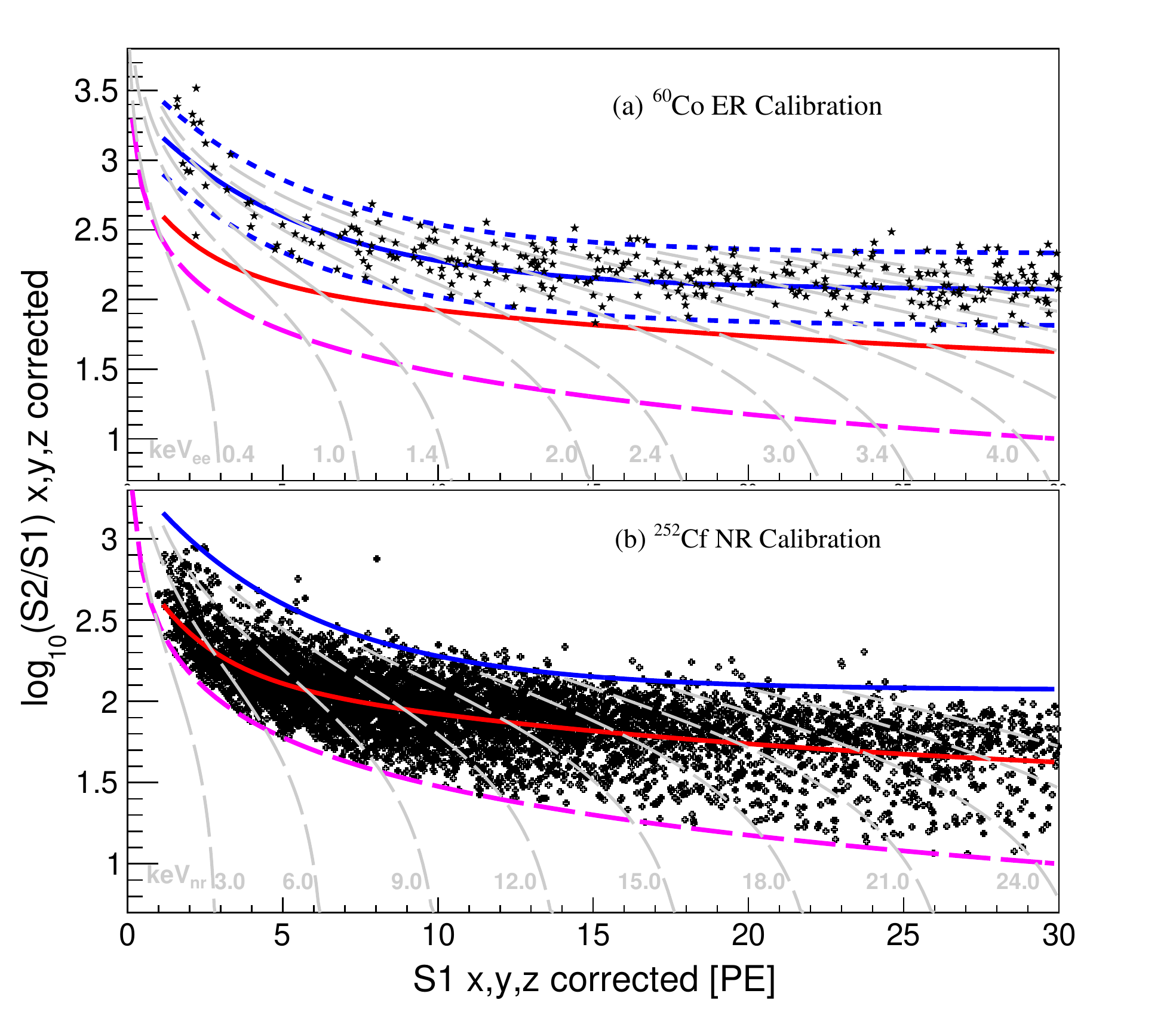}
  \caption{The $\log_{10}(S2/S1)$ versus S1 for (a) ER and (b) NR
  calibration data with means (solid blue and red lines, respectively) 
  and $\pm2\sigma$ ER contours (dashed blue lines). The dashed magenta 
  curve represents the 300\,PE bottom S2 cut. The gray dashed lines are the
  constant energy contours using the combined energy scale based on NEST 
  and the measured PDE and EEE.
  }
  \label{fig:ER_NR_band}
\end{figure}

The detector response to ER and NR events is studied with $^{60}$Co
and $^{252}$Cf calibration data. 
Single scattering events are selected and a fiducial cut of 20\,cm radius and 20--80\,$\mu$s drift time is applied to the reconstructed vertices. 
We select low-energy calibration events with S1 between 2 and 30\,PE,  
and S2 from the bottom 
PMT array greater than 300 PE. The NR purity from the $^{252}$Cf data within this energy range is expected to be approximately 98\% based on a Monte Carlo (MC) simulation. 
The bands from ER and NR calibration data are shown in Fig.~\ref{fig:ER_NR_band}. For the ER calibration band, a total of 278 events are found during 135.4 live-hours of $^{60}$Co calibration data. The mean of the ER band is computed for every 1-PE slice of S1, and fitted with a double exponential form. The width of the band was assumed independent in S1, due to limited statistics. An average ER Gaussian leakage fraction of 0.32\% below the mean of the NR band (or 99.7\% ER rejection efficiency) is obtained based on the width of the ER band. We have performed alternative fittings of the ER band, which generate about 0.1\% change in the leakage fraction. 

\begin{figure}[!htbp]
  \centering
  \includegraphics[width=0.49\textwidth]{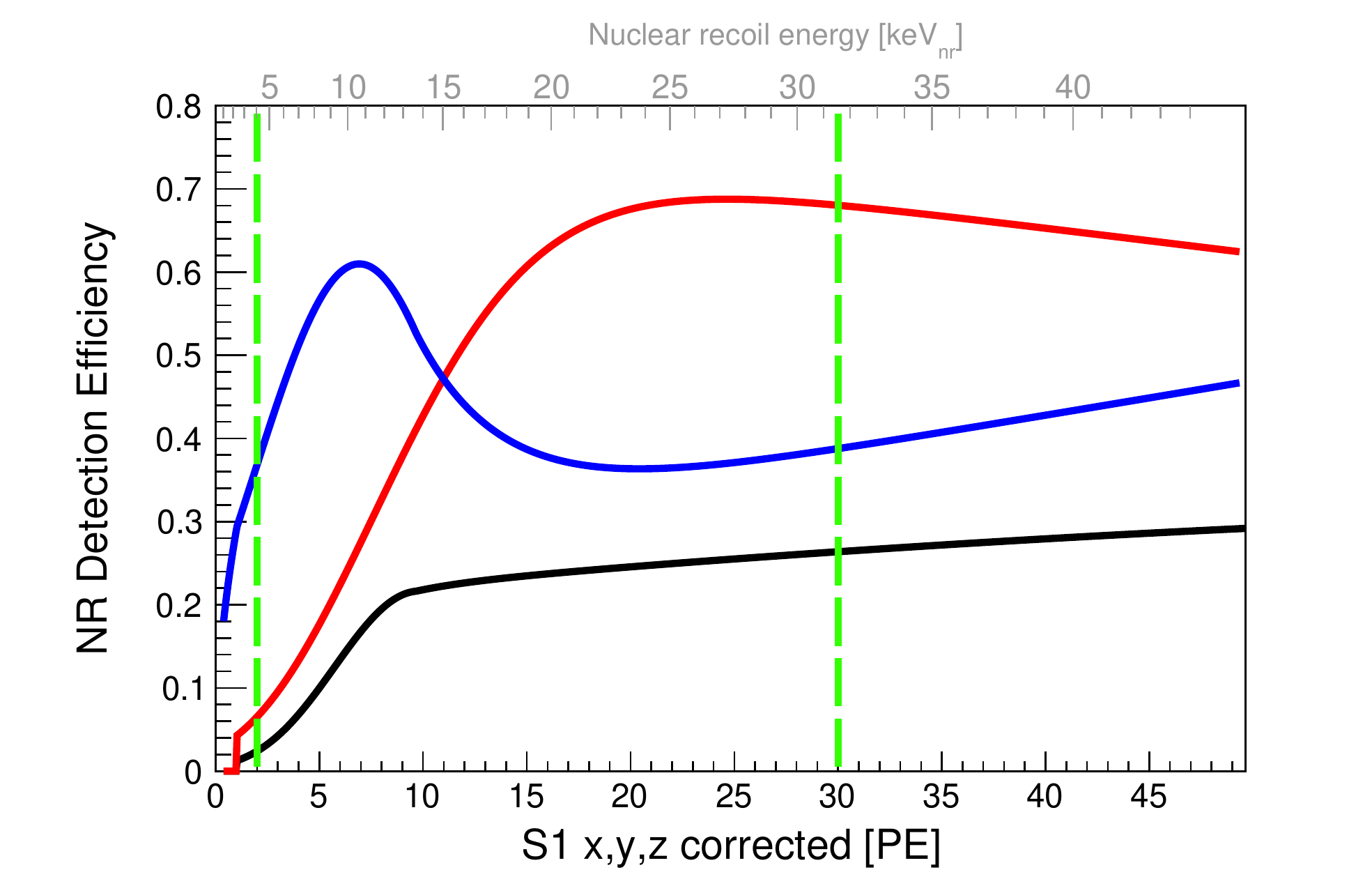}
  \caption{Nuclear recoil detection efficiency  as a function of S1 (the corresponding mean nuclear recoil
    energy are indicated as the ticks on the top). The red, blue, and black curves are the
 	cut efficiency $\epsilon_{\rm cut}$, nuclear recoil acceptance $\mathcal{A}$, and the
    overall NR detection efficiency, respectively.
  }
  \label{fig:efficiency}
\end{figure}

The overall event detection efficiency is the combination of cut efficiency
$\epsilon_{\rm cut}$ and signal acceptance $\mathcal{A}$ of the NR signal window.
The cut efficiency $\epsilon_{\rm cut}$ includes the identification and quality cut efficiencies
on both S1 and S2, and the trigger efficiency on S2 signals. The S1 identification and quality cut efficiencies are evaluated using the low-intensity LED data. The S2 identification efficiency is determined by regrouping closely-packed multiple S1 signals as mis-identified small S2 signals, and is close to 97\%. The S2 quality cut efficiency due to bottom-to-top-charge-ratio cut is
evaluated by selecting events well located on the NR band and
taking the ratio of events with and without the cut. Finally, the S2 trigger efficiency is 
obtained by fitting the measured S2 NR spectrum assuming an exponential form 
of the true energy spectrum. Combining these, $\epsilon_{\rm cut}$ is found as 
a rising function of S1 with a maximum of about 70\% at 25 PE, and then falls
gradually due to the pulse height to area ratio quality cut on the S1 waveform.
Consistent cut efficiency, depicted in Fig.~\ref{fig:efficiency}, is obtained by taking the ratio of the measured NR spectrum to the expected spectrum from the MC. This cut efficiency includes the contribution due to the 300 PE bottom S2 cut. 
The corresponding signal acceptance $\mathcal{A}$ is defined as the ratio of the number of NR events below the mean (i.e. average) of the NR band to the total, also shown in Fig.~\ref{fig:efficiency} together with the overall efficiency. 
The change of the acceptance as a function of S1 indicates the variation of the $\log_{10}(S2/S1)$ distribution in different S1 slices. 
The events on the NR band with suppressed S2 could be due to multiple-scattered neutrons that deposit partial energy below the cathode. Using double-scatter neutrons with the second scatter in the very bottom layer of the TPC as a proxy, we estimate that those below-cathode events could lead to a maximum fractional reduction of 25\% to the overall NR detection efficiency. A detailed study of this will appear in a separate publication.

The analysis results reported here are from 17.4 live days of
DM search data,  taken from May 26 to July 5, 2014. Event rates are summarized in
Table~\ref{tab:dm_rates} for various cut levels. The reduction of background due to single S2 cut is consistent with the MC expectation.
Dark matter candidates are selected by employing identical selection cuts
used in the calibration data.
The signal window for S1 between 2 and 30\,PE
corresponds to a mean energy range of 0.5 to 5.5\,keV$_{\rm ee}$ or
4.1 to 31.6\,keV$_{\rm nr}$ (nuclear recoil) based on the NEST model.
\begin{table}[!htbp]
  \centering
  \begin{tabular}{lrc}
    Cut & \# events & rate (Hz)\\\hline
    all triggers & 4,062,609 & 2.70 \\
    quality cuts & 1,877,707 & 1.25 \\ 
    single-site cut & 1,195,119 & 0.80 \\
    S1 range (2 -- 30\,PE) & 10,268 & 6.83$\times10^{-3}$ \\
    S2$_{\rm bottom}$ range (300 -- 20,000\,PE) & 7,638 & 5.08$\times10^{-3}$ \\
    fiducial volume & 46 &3.06$\times10^{-5}$ \\\hline
  \end{tabular}
  \caption{The event rate of the dark matter data for different cut levels.}
  \label{tab:dm_rates}
\end{table}

\begin{figure}[!htbp]
  \centering
  \includegraphics[width=0.49\textwidth]{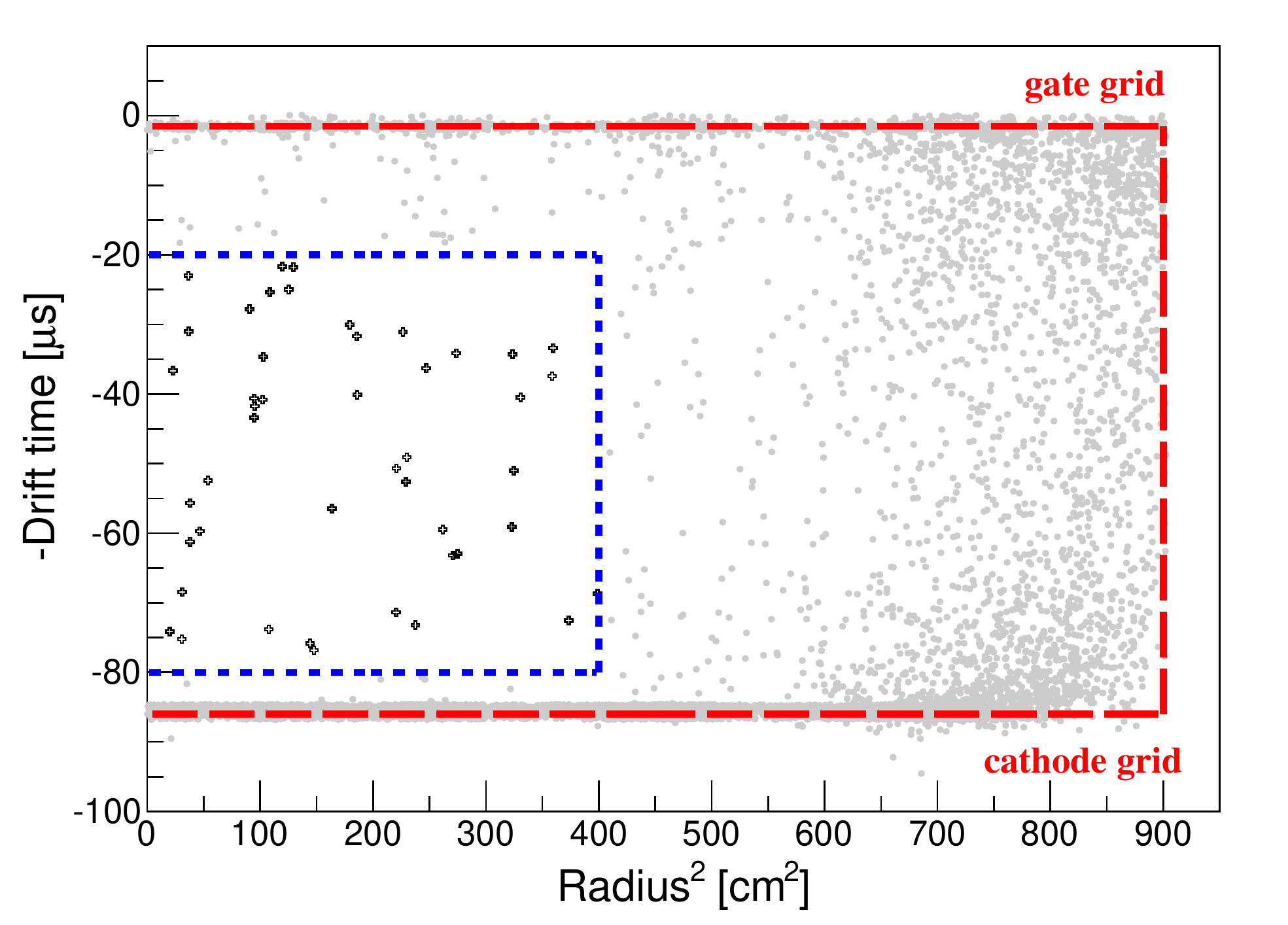}
  \caption{Vertex distribution of all events before the fiducial and
    ER rejection cuts during 17.4 live days of dark matter search data. The 37-kg fiducial volume is contained within the blue dashed box. The location of the detector wall, the gate grid, and the cathode are also indicated in red.}
  \label{fig:vertex_distri_DM}
\end{figure}

The signal vertex distribution before fiducial and ER rejection cuts is
displayed in Fig.~\ref{fig:vertex_distri_DM}. 
The fiducial cut 
indicated by the dashed box is asymmetric in the vertical direction
to provide balanced shielding from radiation originating in the top and bottom PMT arrays.
The cut in $r^2$ was optimized to reduce residual background near the detector walls.
The LXe contained in the fiducial volume is $37.0\pm2.2$\,kg, where the
uncertainty is estimated from the difference between the CoG and NN reconstructions 
for 40\,keV inelastic events in the neutron calibration data.

The measured energy distribution for events in the fiducial volume, after correcting for the detection efficiency, is in agreement with a GEANT4-based MC prediction~\cite{ref:G4}, 
taking into account known background from detector material radioactivities.  
The average background rate of 32$\pm$5 mDRU (DRU = events/keVee/kg/day) 
is consistent with the 43$\pm$11 mDRU MC expectation (Table~\ref{tab:bkg_rates}).
The Kr level is estimated from the delayed coincidence signals from $^{85}$Kr and $^{85m}$Rb decays in the 120-kg target mass to be less than 83 ppt mol/mol (90\% C.L.), assuming an abundance of $^{85}$Kr of $2\times10^{-11}$ in Kr. This is consistent with a 
direct measurement of the xenon sample using a specialized residual gas analysis system with 
a cold trap~\cite{ref:hall}. $^{222}$Rn and $^{220}$Rn in the detector
are identified by their characteristic $\beta$-$\alpha$ delayed coincidences.
The total decay rate in the FV, dominated by $^{222}$Rn, is measured to 
be 0.83$\pm$0.59 mBq, where the uncertainty is dominated by that in the efficiency of the $\beta$-$\alpha$ selection cuts.
This results in a background of 2.7$\pm$2.0 mDRU based on a MC simulation. Gamma events which multiple-scatter in the detector with a large fraction energy deposition in the LXe below the cathode can fake NR events because their ionization energy is only partially captured in the S2 signal. These events, known as the ``gamma-X'' events~\cite{xenon10_0}, contribute  
0.2 events in the 17 days of dark matter data, as estimated by a 
MC simulation.

\begin{table}[!htbp]
  \centering
  \begin{tabular}{lc}
    Source & background level (mDRU) \\\hline
    Top PMT array & 10.9$\pm$1.8 \\
    Bottom PMT array & 4.0$\pm$0.6 \\
    Inner vessel components & 18.5$\pm$10.1 \\
    TPC components & 2.3$\pm$0.8\\
    $^{85}$Kr & $<$3.3 \\
    $^{222}$Rn and $^{220}$Rn & 2.7$\pm$2.0\\
    Outer vessel & 1.3$\pm$0.9 \\\hline
    Total expected & 43$\pm$11\\\hline
    Total observed & 32$\pm$5\\\hline
  \end{tabular}
  \caption{The expected and observed background rates in the fiducial volume. Uncertainties in the MC prediction originate from uncertainties in the material radioactivity screening, except those for Rn and Kr that are due to the 
  uncertainty in the PandaX data. }
  \label{tab:bkg_rates}
\end{table}

\begin{figure}[!htbp]
  \centering
  \includegraphics[width=0.49\textwidth]{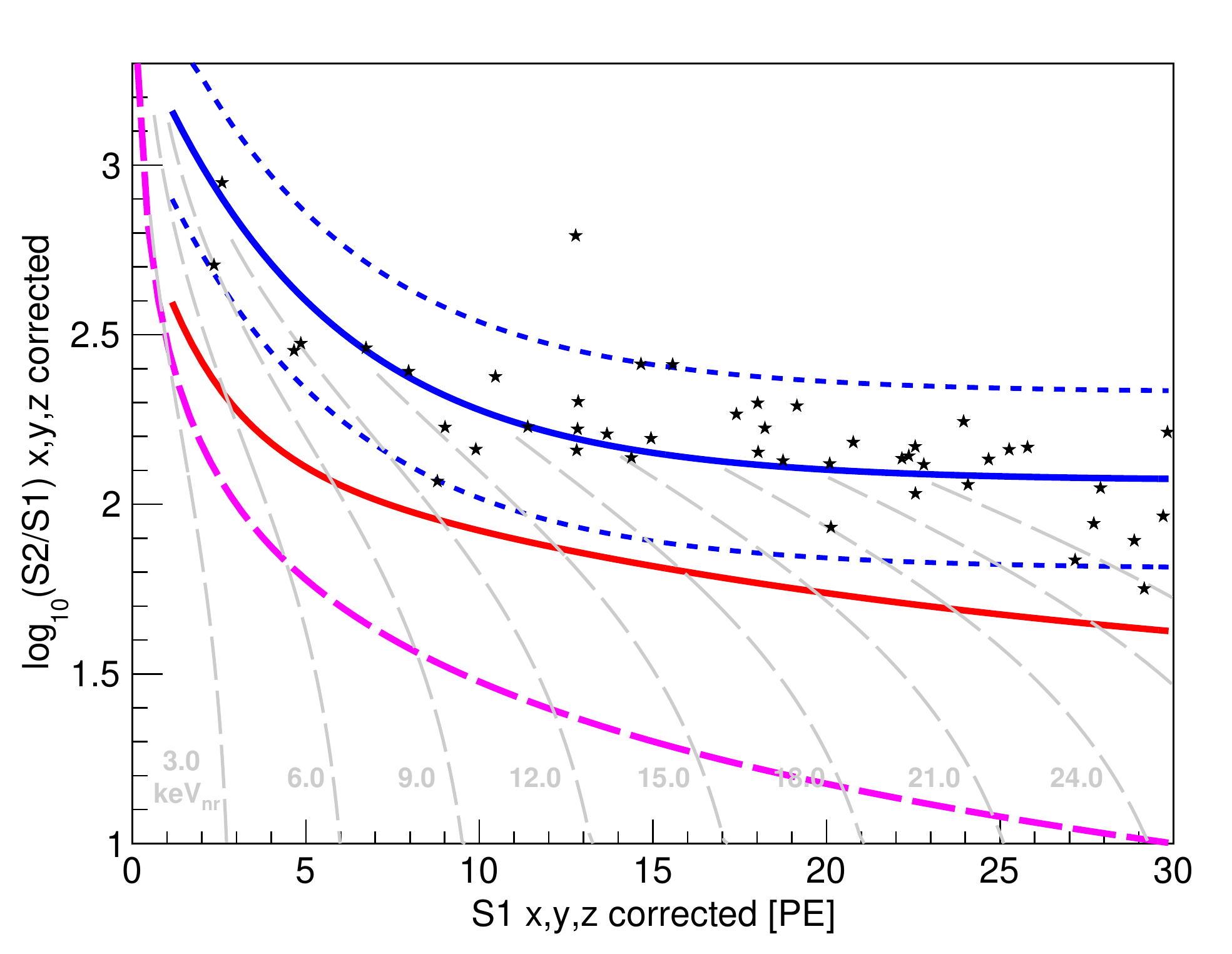}
  \caption{The $\log_{10}(S2/S1)$ versus S1 distribution of events in the fiducial volume from 
	DM search data. No event lies in the signal region. The curves are the same as those defined in   Fig.~\ref{fig:ER_NR_band}}
  \label{fig:dm_band}
\end{figure}

The $\log_{10}(S2/S1)$ versus S1 band from the DM data is shown in Fig.~\ref{fig:dm_band}. No candidate event survives the ER rejection cut. To determine the spin-independent isoscalar WIMP-nucleon scattering cross section as a function of WIMP mass, the WIMP event rate is calculated based on the standard isothermal halo model~\cite{Smith:2007,Savage:2006} with a DM density of 0.3\,GeV/$c^2$/cm$^3$, a local circular velocity of 220\,km/s, a galactic escape velocity of 544 km/s, and an average earth velocity of 245\,km/s. After modeling  
detection efficiencies, Poisson fluctuation in the S1 signal is applied to smear the S1 acceptance. The 90\% C.L. upper limit of the DM signal
is calculated from the Feldman-Cousins statistical model~\cite{ref:fc} 
with no observed event and 
an expected ER Gaussian leakage background of 0.15 event. For a more 
conservative DM limit, we did not add the “gamma-X” estimate into our 
expected background.
The lowest cross section obtained is $3.7\times10^{-44}$
cm$^2$ at a WIMP mass of 49\,GeV/c$^2$.

\begin{figure*}[!htbp]
\begin{centering}
\includegraphics[width=0.95\textwidth]{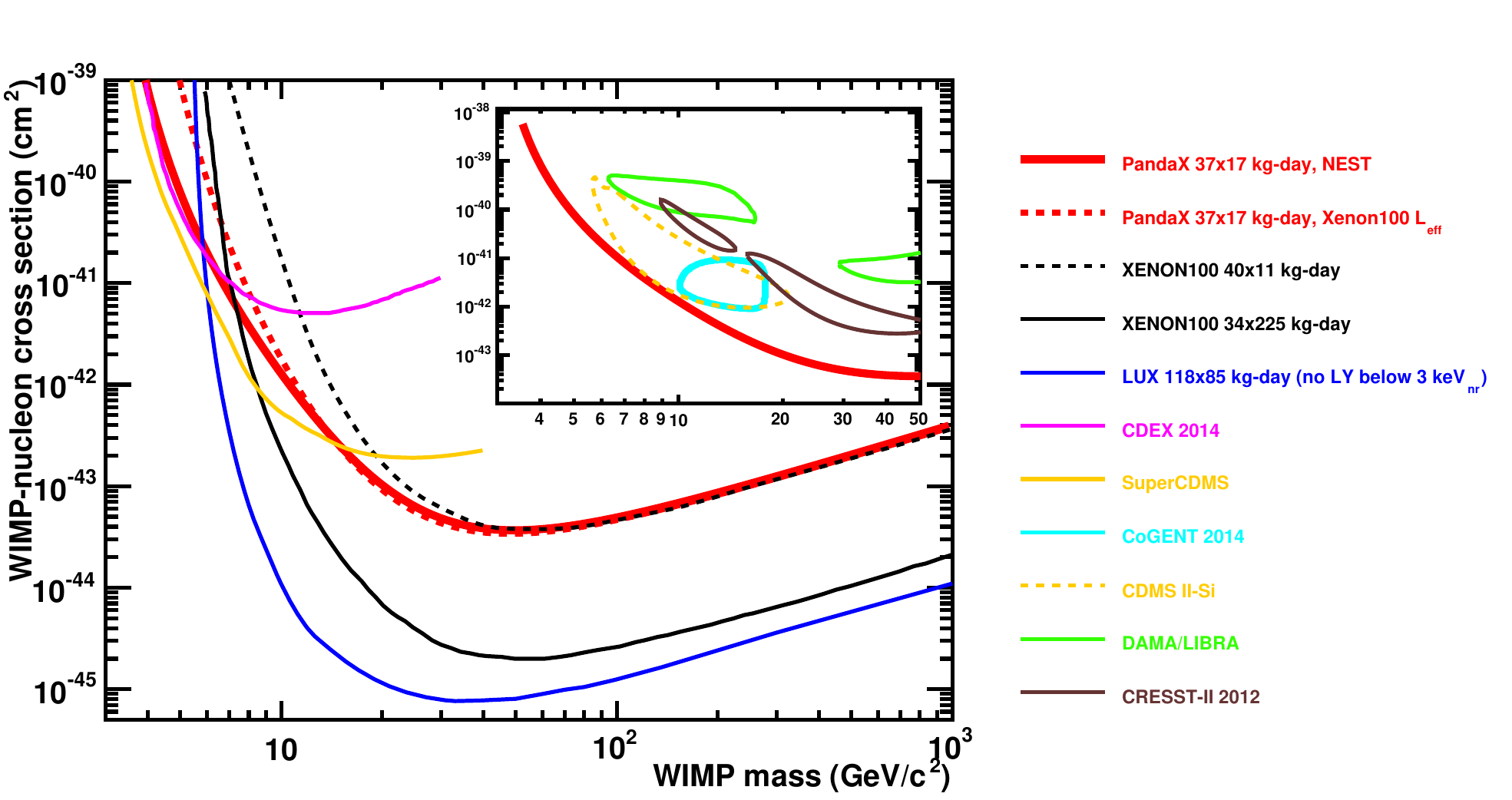}
\par\end{centering}
\caption{\label{fig:sens_final}
  The 90\% C.L. upper limit for spin-independent isoscalar WIMP-nucleon cross section for the PandaX-I experiment (red curves): PandaX-I using $E_{\rm nr}$ and S1 mapping from NEST~\cite{ref:NEST} (red solid) and using $L_{\rm eff}$ from Ref.~\cite{xenon100_first} (red dashed). Recent world results are plotted for comparison: XENON100 first results~\cite{xenon100_first} (black dashed), XENON100 225 day results~\cite{xenon100_final} (black solid),  LUX first results~\cite{ref:lux} (blue),   CDEX 2014 results~\cite{ref:cdex} (magenta), SuperCDMS results~\cite{ref:supercdms} (orange solid), DAMA/LIBRA results~\cite{ref:dama} (green), CoGENT results~\cite{ref:cogent} (cyan), CDMS II-Si results~\cite{ref:cdms-si} (orange dashed), and CRESST-II 2012 results~\cite{ref:cresst1} (brown).
}
\end{figure*}

In Fig.~\ref{fig:sens_final} our results are presented together with
recent world direct detection data~\cite{xenon100_first,xenon100_final,ref:lux,ref:cdex,ref:supercdms,ref:cogent,ref:cresst1,ref:cdms-si,ref:dama}. To quantify the impact of uncertainties in the energy scale on the experimental limit, the calculation is performed using two different $L_{\rm eff}$ (\cite{xenon100_first}) scalings between S1 and $E_{\rm nr}$. The first $L_{\rm eff}$ is taken from the NEST model using the measured PDE of 10.5\%. The second is the conservative
$L_{\rm eff}$ used by XENON100~\cite{xenon100_first} with our 
measured $L_y^{122}$. Below 10\,GeV/$c^2$, the latter gives a more conservative
limit. 
Note that our results show a nominally 
better limit below 6\,GeV/$c^2$ than that from LUX due to that LUX used 
an energy scale with zero light yield below 
3 $\rm keV_{\rm nr}$, which is very conservative 
compared to NEST or other phenomenological 
models (e.g. Ref~\cite{ref:Ji_Leff}). 
Our result is comparable in the high WIMP mass region to that of Ref. \cite{xenon100_first}
with similar exposure, and is significantly 
more constraining in the low-mass region, demonstrating the advantage of the low-energy 
threshold of the PandaX-I detector.  
At the 90\% C.L., our results are incompatible with 
the spin-independent isoscalar WIMP interpretation of previously reported
observed signals from DAMA, CoGeNT, CRESST and CDMS II-Si~\cite{ref:dama, ref:cogent, ref:cresst1, ref:cdms-si}.
In the high WIMP mass region, our result confirms the power of the LXe
dual-phase technique as one of the leading technologies to probe the 
theoretically-favored DM particles, e.g., predicted by supersymmetric models. 

In summary, we report the first results using 17.4 live-days of data in the 37-kg fiducial mass from the PandaX-I dark matter experiment at CJPL.
They place strong constraints in the low WIMP mass region which are being actively studied by many other experiments. 
PandaX-I continues to take data and explore low-mass WIMPs. PandaX-II, with about ten times larger fiducial mass and improved background properties, is
being constructed to reach a sensitivity beyond the current best limits 
in a wide WIMP mass range.

This project has been supported by a 985-III grant from Shanghai Jiao Tong University, a
973 grant from Ministry of Science and Technology of China (No. 2010CB833005), a grant
from National Science Foundation of China (No.11055003), and a grant from the Office
of Science and Technology in Shanghai Municipal Government (No. 11DZ2260700). Xun Chen acknowledges support from China Postdoctoral Science Foundation Grant 2014M551395.
The project has also been sponsored by Shandong University, Peking University, the University
of Maryland, and the University of Michigan. We would like to thank many people including
Elena Aprile, XianFeng Chen, Carter Hall, T. D. Lee, ZhongQin Lin, Chuan Liu, Lv Lv, 
YingHong Peng, WeiLian Tong, HanGuo Wang, James White, YueLiang Wu, QingHao Ye,  
Qian Yue, and HaiYing Zhao for help and discussion at
various level. We are particularly indebted to Jie Zhang for his strong support and crucial
help during many stages of this project. Finally, we thank the following organizations and
personnel for indispensable logistics and other supports: the CJPL administration including
directors JianPing Cheng and KeJun Kang and manager JianMin Li, Yalong River Hydropower
Development Company, Ltd. including the chairman of the board HuiSheng Wang, and manager
XianTao Chen and his JinPing tunnel management team from the 21st Bureau of the China
Railway Construction Co.

\end{document}